\documentclass[aps,pra,superscriptaddress,showpacs]{revtex4} 
 
\usepackage{dcolumn}  
\usepackage{graphicx}  
\usepackage{epsf}  
\usepackage{epsfig}  
\usepackage{amsmath}  
\usepackage{bm}  
\usepackage{times}  
\usepackage{nicefrac}  
\usepackage{color}  

\newcolumntype{.}{D{x}{}{-1}}  
  
\def\corresponds{{\lower.2ex\hbox{=}}{\rm\kern-.75em^\triangle}}  
\def\succsim{\succ\kern-.9em_\sim\kern.3em}  
\def\precsim{\prec\kern-1em_\sim\kern.3em}  
\def\slantfrac#1#2{\kern1em^{#1}\kern-.3em/\kern-.1em_{#2}}  
\def\lfrac#1#2{{}^{#1\!}\kern-.0em/_{#2}}  
  
\def\buildrel#1\under#2{\mathrel{\mathop{\kern0pt #2}\limits_{#1}}}

\newcommand{\hf} {\frac{1}{2}}

\def\mr#1{{\mathrm{#1}}}

\begin{document}  
\preprint{Version 1.0}  
 
\title{Applicability of layered sine-Gordon models to  
layered superconductors: \\ II. The case of magnetic coupling} 
 
\author{I. N\'andori}  
\affiliation{Institute of Nuclear Research, P.O.Box 51,  
H-4001 Debrecen, Hungary}  
\affiliation{Max--Planck--Institut f\"ur Kernphysik,  
Postfach 103980, 69029 Heidelberg, Germany} 
 
\author{K. Vad}  
\affiliation{Institute of Nuclear Research, P.O.Box 51,  
H-4001 Debrecen, Hungary}  
  
\author{S. M\'esz\'aros}  
\affiliation{Institute of Nuclear Research, P.O.Box 51,  
H-4001 Debrecen, Hungary}  
 
\author{U. D. Jentschura}  
\affiliation{Max--Planck--Institut f\"ur Kernphysik,  
Postfach 103980, 69029 Heidelberg, Germany}  
\affiliation{National Institute of Standards and Technology,   
Mail Stop 8401, Gaithersburg, MD20899-8401, USA}  
  
\author{S. Nagy}  
\affiliation{Department of Theoretical Physics, University of Debrecen,  
Debrecen, Hungary}  
  
\author{K. Sailer}  
\affiliation{Department of Theoretical Physics, University of Debrecen,  
Debrecen, Hungary}

\begin{abstract}  
In this paper, we propose a quantum field theoretical renormalization  
group approach to the vortex dynamics of magnetically coupled layered  
superconductors, to supplement our earlier investigations on  
the Josephson-coupled case.  
We construct a two-dimensional multi-layer sine-Gordon  
type model which we map onto a gas of topological  
excitations. With a special choice of the mass matrix  
for our field theoretical model,  
vortex dominated properties of magnetically coupled  
layered superconductors can be described. 
The well known interaction potentials of fractional flux 
vortices are consistently obtained from our field-theoretical analysis, 
and the physical parameters (vortex fugacity and temperature parameter)  
are also identified. We analyse the phase structure of the multi-layer  
sine--Gordon model by a differential renormalization group method  
for the magnetically coupled case 
from first principles. The dependence of  
the transition temperature on the number of layers is found to be  
in agreement with known results based on other methods.  
\end{abstract}  
  

\maketitle  
 
% 
% INTRODUCTION 
% 
\section{INTRODUCTION} 
 
Recently, we have shown that layered sine-Gordon type models  
are probably not suitable for the description of Josephson-coupled  
layered superconductors, because the linear, confining potential  
that binds the vortices together cannot be obtained from the  
interaction of the topological excitations of the model, no matter 
how the interlayer interaction term is chosen~\cite{NaEtAl2007jpc}. 
On the other hand, vortex dominated properties of high $T_{\rm c}$  
layered superconductors and  
other types of layered materials, e.g. superconducting  
sandwiches, have already received a considerable amount of  
attention (see, e.g., Refs.~\cite{Pe1964,dG1966,Ef1979, 
ArKr1990,BuFe1990,FeGeLa1990,Cl1991,Fi1991,KoVi1991,Pu1993, 
BlEtAl1994,MiKoCl2000,ClemPancake,GoHo2005,CoGeBl2005}), and the  
intuitively obvious connection of sine-Gordon models to these 
materials makes one wonder if at least, a restricted applicability 
of the layered, field-theoretical model persists. 
We also observe that recently, there is an increasing interest  
in the literature 
\cite{BeCaGi2007,BeCaGi2007magnetic,Ar2007}  
in constructing sine--Gordon type field theoretical models in  
order to understand better the vortex dynamics in layered  
superconducting systems. Our aim in this paper to follow this  
route by constructing a two-dimensional multi-layer sine-Gordon  
type model which can be used to describe the vortex behaviour of  
magnetically as opposed to Josephson-coupled layered superconductors,  
and to contrast and enhance our recent investigation~\cite{NaEtAl2007jpc}.  
 
In a two-dimensional (2D) isolated superconducting thin film, the  
Pearl vortices~\cite{Pe1964,dG1966} are naturally identified as  
the topological excitations and can be considered as the charged  
analogues of the vortices in a 2D superfluid which generate the  
Kosterlitz--Thouless--Berezinski (KTB) phase transition~\cite{KTBPhase}.  
The logarithmic interaction between the vortices of the superfluid  
extends to infinity and as a consequence they remain bound below  
the finite KTB transition temperature ($T^{\star}_{\rm KTB}$) and  
dissociate above it~\cite{KTBPhase}. Since the Pearl vortices carry  
electric charge, they always remain unbound due to the screening length  
$\lambda_{\rm eff}$ generated by the electromagnetic field which cuts off  
the logarithmic interaction \cite{PiVa2000,BlEtAl1994,ReEtAl1996} and  
leads to the absence of any KTB phase transition. However, for realistic  
finite 2D superconducting films where the lateral dimension of the  
film can be smaller then the screening length $R_0 < \lambda_{\rm eff}$  
the KTB transition can be restored \cite{PiVa2000,BlEtAl1994}.  
This constitutes an intrinsic finite size effect.  
 
In layered materials, the interlayer coupling modifies the 2D picture 
and leads to new types of topological defects. If the layers are  
coupled by Josephson coupling (like for many HTSC materials), the  
vortex-antivortex pairs on the same layer interact with each other  
via a logarithmic term for small distances but they feel a linear  
confining potential for large distances (see e.g. \cite{BlEtAl1994}  
and references therein). The vortices in neighboring layers always  
interact via a linear potential which can couple them by forming  
vortex loops, rings, or vortex ``strings'' piercing all layers.  
 
If the layers are coupled by purely magnetic interaction (e.g. in  
artificially produced superlattices where the Cooper pair tunneling  
between the superconducting layers is suppressed by relatively large  
insulating layers) the topological defects for a system which consists  
of infinitely many layers are pancake vortices~\cite{Cl1991,ClemPancake}  
which undergo a KTB phase transition at $T^{\star}_{\rm KTB}$. 
As explained e.g.~in Ref.~\cite{CoGeBl2005}, the  
Josephson coupling can be essentially neglected when 
the confinement length, i.e.~the length scale at  
which the linear confining potential due to the 
Josephson coupling dominates over the logarithmic interaction 
due to magnetic effects, is pushed beyond the effective 
screening length for the logarithmic interaction among vortices.  
This situation is present when the tunneling between the  
superconducting layers is suppressed by relatively large insulating  
layers, and a proposal for a experimental 
realization has recently been given~\cite{CoGeBl2005}. 
For a finite number $N$ of magnetically coupled layers, the Pearl type  
vortex stack~\cite{Pe1964} is broken up into a number of coupled pancake  
vortices of fractional flux~\cite{Pu1993,MiKoCl2000,ClemPancake}, and this  
configuration undergoes a KTB-type phase transition at a layer-dependent  
temperature $T^{(N)}_{\rm KTB}=T^{\star}_{\rm KTB}(1-N^{-1})$ which is  
connected with the dissociation of the stack. This result has been  
obtained on the basis of the entropy method first introduced in the  
ground-breaking work~\cite{Pu1993}. Recently, a real space  
renormalization-group (RG) analysis of the case $N=2$ has been performed  
in Ref.~\cite{CoGeBl2005} using the dilute gas approximation.  
A priori,it appears to be rather difficult to generalize this RG  
analysis for $N>2$ layers.  
 
In general, the Ginzburg--Landau (GL) theory \cite{Gi1952} provides  
us with a good theoretical framework in which to investigate the 
vortex dynamics in  
thin films and in layered materials. Several equivalent models, like  
field-theoretical, statistical spin models and a gas of topological 
defects have also been used to consider the vortex properties  
of films and layered systems. The 2D-GL, 2D-XY and the 2D Coulomb  
gas models (see e.g. \cite{BlEtAl1994,NaEtAl2007jpc} and references  
therein) are considered as the appropriate theoretical background for  
the vortex dynamics of superfluid films. The field theoretical  
counterpart is the 2D sine-Gordon (SG) model \cite{SG2D}. Both kinds   
of these models belong to the same universality class and produce 
the KTB phase transition. For superconducting films one has to 
consider the 2D-GL model in the presence of electromagnetic  
interactions \cite{BlEtAl1994} or the equivalent gas of topological  
excitations, the 2D Yukawa gas \cite{PiVa2000}. The corresponding  
field theory is the 2D-SG model with an explicit mass term, the  
massive 2D-SG model \cite{PiVa2000}.  
 
For Josephson-coupled layered superconductors in the case of very  
large anisotropy one should investigate the layered GL model  
including the Josephson coupling between the layers~\cite{BlEtAl1994}  
(i.e. the Lawrence-Doniach model~\cite{LaDo1971}). In case of not  
too large anisotropy on can use the anisotropic, continuous GL  
theory~\cite{Gi1952,BlEtAl1994,ChDuGu1995} which can be mapped  
onto the isotropic GL model by an appropriate rescaling  
method~\cite{RESCALE}. The corresponding spin model is the  
3D-XY model~\cite{3DXY} and the equivalent gases of topological  
excitations are the layered vortex~\cite{Pi1995prb} or  
vortex-loop~\cite{3DXY} gases. There are attempts in the literature  
to construct the field theoretical countpart of the isotropic   
model~\cite{Sa1978}. In case of strong anisotropy, the layered  
sine--Gordon (SG) model~\cite{LSGPierson} has been proposed as a  
candidate model where the interlayer interaction between the  
topological defects has been described by a mass matrix  
which couples the SG fields  
\begin{align}  
\hf \underline{\varphi}^{\rm T} \, \underline{\underline m}^2   
\, \underline{\varphi} \equiv  
\sum_{n = 1}^{N-1} \frac{J}{2} (\varphi_{n+1}-\varphi_n)^2  
\nonumber 
\end{align}  
where $\underline{\varphi}=\left(\varphi_1, \dots, \varphi_{N}\right)$ 
and $\varphi_n$ (n=1,...,N) are one component scalar fields. 
Recently, we showed in Ref.~\cite{NaEtAl2007jpc} that the layered SG  
model with the above mass matrix is not apropriate for the description 
of vortex dynamics of Josephson coupled layered superconductors.  
 
In case of purely magnetically coupled layered systems, the  
layered GL model has to be used but excluding the Josephson  
coupling. Although the interaction potentials between the topological  
defects of magnetically coupled layered systems are given in  
Refs.~\cite{CoGeBl2005,ClemPancake,BuFe1990,KoVi1991}, no field  
theoretical model has been proposed for the description of vortex  
dynamics in a finite system of magnetically coupled superconductors.  
 
Here, our aim is to open a new platform for considering the vortex  
dynamics of magnetically coupled layered systems by constructing a  
multi-layer sine--Gordon (MLSG) type field theoretical model where the  
two-dimensional sine--Gordon (2DSG) fields  characterizing the layers 
are coupled by an appropriate general mass matrix,  
\begin{align}   
\hf \underline{\varphi}^{\rm T} \, {\underline{\underline M}}^2  
\underline{\varphi} \equiv  
\hf G \left(\sum_{n=1}^N \varphi_n \right)^2 \,. 
\nonumber 
\end{align}  
By the exact mapping of the MLSG model onto an equivalent gas of  
topological defects, we recover the interaction potential given in  
Refs.~\cite{BuFe1990,KoVi1991,ClemPancake,CoGeBl2005}  
and, hence, prove the applicability of the model. We analyse the phase  
structure of the MLSG model by a differential renormalization group  
(RG) method performed in momentum space, which is in general easier  
to perform than that in real space, and determine the layer-dependence of  
$T^{(N)}_{\rm KTB}$. In our field theoretical RG approach, the RG  
flow can be calculated in one step for an arbitrary number of layers,  
and the study of the intrinsic finite size effect of thin film  
superconductors \cite{BlEtAl1994,PiVa2000} and of finite layered  
systems is facilitated. 
 
This paper is organized as follows. In Sec.~\ref{sec2}, we define 
the multi-layer sine--Gordon model and show by its exact mapping 
onto the equivalent gas of topological excitations that it is 
suitable to describe the vortex dominated properties of magnetically 
coupled layered superconductors. In Sec.~\ref{sec3}, a renormalization 
group analysis of the multi-layer sine--Gordon model is 
performed within the framework of the Wegner--Houghton renormalization  
group method, in momentum space for general $N$, 
and with a solution that spans the entire domain from the  
ultraviolet (UV) to the infrared (IR). 
The layer-number dependence of the critical temperature  
of the multi-layer sine--Gordon model is determined by using the 
mass-corrected linearized RG flow. Conclusions are reserved  
for Sec.~\ref{sec4}.  
 
% 
% Multi-Layer Sine--Gordon Model 
% 
\section{Multi-Layer Sine--Gordon Model} 
\label{sec2} 
 
The multi-layer sine--Gordon (MLSG) model consists of $N$ coupled  
two-dimensional sine--Gordon (2D-SG) models of identical ``frequency'' 
$b$, each of which  
corresponds to a single layer described by the scalar fields  
$\varphi_n$ $(n=1,2,\ldots,N)$. Its Euclidean bare action  
(we imply here the sum over $\mu = 1,2$) 
\begin{align}  
\label{mlsg}  
S[\underline{\varphi}] = \int {\rm d}^2 r  
\biggl[ \hf (\partial_{\mu} \underline\varphi)^{\rm T}  
(\partial_{\mu} \underline\varphi) + 
V( \underline\varphi) \biggr]  
\end{align}  
contains the interaction terms   
\begin{align}  
V(\underline{\varphi}) =  
\hf \underline\varphi^{\rm T} \, {\underline{\underline M}}^2   
\underline\varphi - \sum_{n=1}^N  y_n \cos (b \, \varphi_n) 
\label{pepot}  
\end{align}  
with the $O(N)$ multiplet $\underline{\varphi}= 
\left(\varphi_{1}, \dots, \varphi_{N}\right)$. 
We can choose the fugacities $y_n > 0$ without loss 
of generality, ensuring that the  
zero-field configuration is a local minimum of the action 
(see Chap.~31 of Ref.~\cite{ZJ1996}). 
The mass-matrix describes the interaction between the layers  
and is chosen here to be of the form 
\begin{align} 
\label{mass_matrix} 
\underline{\varphi}^{\rm T} \, {\underline{\underline M}}^2  
\underline{\varphi} 
= G \left(\sum_{n=1}^N a_n \varphi_n \right)^2 \,, 
\end{align} 
where $G$ is the strength of the interlayer interactions, 
and the $a_n$ are free parameters.  
As will be explained below, any choice with $a^2_n = 1$ for all $n=1,\ldots,N$ 
reproduces exactly the same layer-dependence of $T^{(N)}_{\rm KTB}$  
as found in Refs.~\cite{Pu1993,CoGeBl2005}.  
In this case, the layers can be assumed 
to be equivalent and, as a consequence, the fugacity $y_n \equiv y$  
for $n=1,2,\ldots,N$.  
The most obvious choice fulfilling $a^2_n = 1$, 
namely $a_n =1$ for all $n=1,\ldots,N$, 
reproduces the interlayer interaction between  
pancake vortices given, e.g., 
in Eq.~(89) of Ref.~\cite{ClemPancake}, 
and we will restrict our attention to this choice in the  
following. 
 
The MLSG model 
has a discrete symmetry under the shift of the field variable 
$\underline{\varphi} \to \underline{\varphi} + \underline{\Delta}$  
with $\underline{\Delta} = \left( l_1 2\pi/b, \dots, l_N 2\pi/b \right)$  
where the ``last'' integer $l_N = -\sum_{n=1}^{N-1} l_n$ is fixed  
but all the other  
integers $l_n$ ($n=1,\ldots,N-1$) can be chosen freely (to see this, 
one just diagonalizes the mass-matrix). 
The single non-vanishing mass eigenvalue is 
$M_N = \sqrt{N G}$, and hence the model  
possesses $N-1$ massless 2D-SG fields and a single massive 2D-SG   
field. After the diagonalization of the mass matrix by a  
suitable rotation of the fields,  
the model thus is invariant under the independent separate shifts of  
$N-1$ massless fields, but the explicit mass term of the single  
massive mode breaks the periodicity in the ``massive'' direction  
of the $N$-dimensional internal space.  
 
One crucial observation is that the partition function of the MLSG model, 
whose path-integral formulation reads 
\begin{equation} 
\label{Z_mlsg} 
{\cal Z} = {\mathcal N} \int {\mathcal D} [\underline{\varphi}]  
\exp{\left(-S[\underline{\varphi}]\right)}, 
\end{equation} 
can be identically rewritten in terms of an equivalent gas of  
topological excitations (vortices), whose interaction potentials 
are exactly equivalent to those of  
Refs.~\cite{BuFe1990,KoVi1991,CoGeBl2005}. 
This finding constitutes a generalization of known connections of the  
$d$-dimensional globally neutral Coulomb gas and the $d$-dimensional 
sine--Gordon model, as discussed in Chap.~32 of Ref.~\cite{ZJ1996}, 
and can be seen as follows. In Eq.~(\ref{mlsg}), one artifically introduces 
the vectors $f_n \equiv \left( \delta_{1n}, \dots, \delta_{Nn} \right)$ 
as projection operators to rewrite $\sum_{n=1}^N \cos(b \, \varphi_n)  
= \sum_{n=1}^N \cos(b \, \underline{f}_n^{\rm T}  \underline{\varphi})$, 
one expands the periodic piece of the partition function  
(\ref{Z_mlsg}) in a Taylor series, and one introduces the integer-valued 
charges $\sigma_\alpha = \pm 1$ of the topological defects which are subject  
to the neutrality condition $\sum_{\alpha=1}^{2\nu} \sigma_\alpha = 0$. 
This leads to the intermediate result, 
\begin{align} 
\label{step1} 
& {\cal Z} = {\mathcal N} \sum_{\nu =0}^\infty \frac{(y/2)^{2\nu}}{(2\nu)!}  
\prod_{i=1}^{2 \nu} \left( \sum_{n_i = 1}^N \int {\rm d}^2 r_i \right) 
\sum_{\begin{array}{c}  
\scriptstyle \sigma_1, \dots, \sigma_{\nu} = \pm 1 \\ 
\scriptstyle \sigma_{\nu+\gamma} = -\sigma_\gamma , \; 
\gamma \in \{ 1, \dots \nu \} \end{array}} 
\\ 
& \times  
\int {\mathcal D}[\underline{\varphi}] 
\exp{\left[-\int\!\! {\rm d}^2 r \, \frac{1}{2}  
\underline\varphi^{\rm T} \,  
(-\partial^2 + {\underline{\underline M}}^2)   
\underline\varphi 
+ {\rm i} \, b \, {\underline\rho}^{\rm T} \, \underline\varphi \right]}, 
\nonumber 
\end{align} 
where $\partial^2 \equiv \partial_\mu \partial_\mu$, 
and  
\begin{equation} 
{\underline \rho}(r) = \sum_{\alpha =1}^{2\nu}  
\sigma_{\alpha} \delta(r - r_{\alpha})  
\underline{f}_{n_{\alpha}}\,. 
\end{equation} 
We have thus placed the $2 \nu$ vortices, labeled by the index $i$, 
onto the $N$ layers, with vortex $i$ being placed onto the layer $n_i$.  
The Gaussian integration in Eq.~(\ref{step1}) can now be performed  
easily, and the inversion of the matrix  
$-\partial^2 + {\underline{\underline M}}^2$ 
can be accomplished by going to momentum space. Via a subsequent 
back-transformation to coordinate space, we finally arrive at the  
result 
\begin{align} 
\label{gte} 
& {\cal Z} =  
\sum_{\nu =0}^\infty \frac{(y/2)^{2\nu}}{(2\nu)!}  
\left(\prod_{i=1}^{2\nu} 
\sum_{n_{i}=1}^N   \int {\rm d}^2 r_i \right)  
\sum_{\begin{array}{c}  
\scriptstyle \sigma_1, \dots, \sigma_{\nu} = \pm 1 \\ 
\scriptstyle \sigma_{\nu+\gamma} = -\sigma_\gamma , \; 
\gamma \in \{ 1, \dots \nu \} \end{array}} 
\\ 
& \exp{\left[-\frac{b^2}{2}  
\sum_{\alpha,\gamma=1}^{2\nu} 
\sigma_{\alpha} \sigma_{\gamma}    
\left( \delta_{n_{\alpha}n_{\gamma}} A_{\alpha \gamma} + 
(1 - \delta_{n_{\alpha}n_{\gamma}}) B_{\alpha \gamma}\right) 
\right]} \,, 
\nonumber 
\end{align} 
where $\delta_{nm}$ represents the Kronecker-delta. 
Equation (\ref{gte}) implies that  
the parameter $b^2$ in Eq.~(\ref{pepot}) can naturally be identified  
as being proportional to the inverse of the temperature of the gas,  
$b^2 \propto T^{-1}$. The potentials  
$A_{\alpha \gamma} \equiv A(\vec{r}_\alpha, \vec{r}_\gamma)$ and  
$B_{\alpha \gamma} \equiv B(\vec{r}_\alpha, \vec{r}_\gamma)$  
are the intralayer and interlayer  
interaction potentials, respectively. They read 
\begin{subequations} 
\label{pot_limit} 
\begin{align} 
\label{A} 
A_{\alpha \, \gamma} =&  -\frac{1}{2\pi} \frac{N-1}{N}  
\ln{\left(\frac{r_{\alpha \gamma}}{a}\right)}  
%\\ 
+ \frac{1}{2\pi} \frac{1}{N}  
\left[K_0\left(\frac{r_{\alpha \gamma}}{\lambda_{\rm eff}}\right)  
- K_0\left(\frac{a}{\lambda_{\rm eff}}\right)\right] 
\nonumber\\[2ex] 
=& \left\{ \begin{array}{cc}  
-\frac{1}{2\pi} \ln\left(\frac{r_{\alpha \gamma}}{a}\right) &  
\quad (r_{\alpha \gamma} \ll \lambda_{\rm eff}) \\[2ex] 
-\frac{1}{2\pi} \left[\frac{N-1}{N} 
\ln\left(\frac{r_{\alpha \gamma}}{\lambda_{\rm eff}}\right) 
-\ln\left(\frac{\lambda_{\rm eff}}{a}\right)\right] & 
\quad (r_{\alpha \gamma} \gg \lambda_{\rm eff}) 
\end{array} \right. 
%\nonumber 
\end{align} 
where $r_{\alpha \gamma} = \vert \vec{r}_{\alpha} - \vec{r}_{\gamma}\vert$, 
and 
\begin{align} 
\label{B} 
B_{\alpha \, \gamma} =& \frac{1}{2\pi}  
\frac{1}{N}   
%\\ 
\left(\ln{\left(\frac{r_{\alpha \gamma}}{a}\right)}  
+ \left[K_0\left(\frac{r_{\alpha \gamma}}{\lambda_{\rm eff}}\right)  
- K_0\left(\frac{a}{\lambda_{\rm eff}}\right)\right]  
\right) 
\nonumber\\[2ex] 
=& \left\{ \begin{array}{cc} 
0 & \quad (r_{\alpha \gamma} \ll \lambda_{\rm eff}) 
\\[2ex] 
\frac{1}{2\pi} \frac{1}{N} 
\ln\left(\frac{r_{\alpha \gamma}}{\lambda_{\rm eff}}\right) & 
\quad (r_{\alpha \gamma} \gg \lambda_{\rm eff}) 
\end{array} \right. . 
%\nonumber 
\end{align} 
\end{subequations} 
$K_0(r)$ stands for the modified Bessel function of the  
second kind, $a$ is the lattice spacing which serves as an UV  
cutoff and an effective screening length $\lambda_{\rm eff}$ is  
introduced which is related inversely to the non-zero mass  
eigenvalue of the mass matrix (\ref{mass_matrix}),  
$\lambda^{-1}_{\rm eff} = M_N = \sqrt{N G}$.  
The relation $K_0(r) = -\ln(r) + \ln 2 -\gamma_{\rm E} + {\cal O}(r)$ 
has been used in the derivation of the asymptotic short- and  
long-range forms in Eqs.~(\ref{A}) and~(\ref{B}), and only 
the leading logarithmic terms are indicated 
($\gamma_{\rm E} = 0.577216\dots$ is Euler's constant). 
 
The interaction potentials (\ref{pot_limit}) have the same asymptotic  
behavior as the vortices of magnetically coupled superconducting  
layers \cite{BuFe1990,KoVi1991,ClemPancake,CoGeBl2005} 
[for the intralayer and interlayer interactions see  
Eqs.~(86) and~(89) of Ref.~\cite{ClemPancake}, under the  
substitution $\Lambda_D =\Lambda_s /N$]. 
This observation shows 
that the MLSG field theory is suitable to describe the vortex dynamics  
in magnetically coupled layered systems. A few remarks are now in order.  
(i) The prefactor $(N-1)/N$ appearing in the intralayer interaction  
indicates the existence of vortices with fractional flux in the MLSG  
model.  
(ii) For small distances $r\ll\lambda_{\rm eff}$, the interlayer  
interaction $B$ disappears and the intralayer potential $A$ has the  
same logarithmic behaviour with full flux as that of the pure 2D-SG  
model (which belongs to the same universality class as the  
2D-XY model and the 2D Coulomb gas). Therefore, the MLSG model for  
small distances behaves as an uncoupled system of 2D-SG models. 
(iii) For the case $N=1$, there exists no interlayer interaction, 
and the intralayer potential is logarithmic for small distances and 
vanishes for large distances.  
Consequently, there are always free, non-interacting vortices  
in the model which push the KTB transition temperature to zero.  
The MLSG model for a single layer reduces to the massive 2D-SG model  
discussed in  
Refs.~\cite{BlEtAl1994,PiVa2000,ReEtAl1996,BeCaGi2007magnetic}  
where 
the periodicity in the internal space is broken and the KTB transition 
is absent.  
(iv) In the bulk limit $N\to\infty$, the effective screening length  
and the interlayer interaction disappear ($\lambda_{\rm eff}\to 0$,  
$B_{\alpha \gamma}\to 0$), and the intralayer potential has a  
logarithmic behaviour with full flux, thus the MLSG model  
predicts the same behaviour as that of the pure 2DSG model  
with $T^{(\infty)}_{\rm KTB} = T^{\star}_{\rm KTB}$.  
Alternatively, one may observe that for $N\to \infty$, 
the effect of the infinitely many zero-mass modes dominates over 
the effect of the single remaining massive mode entirely, 
leading to a constant limit for the transition temperature 
as $N \to \infty$.  
 
For $N=2$ layers, the MLSG model [with the choice $a_n = (-1)^{n+1}$] 
has been proposed to describe the 
vortex properties of Josephson coupled layered superconductors  
\cite{LSGPierson}. However, the above discussed mapping indicates that  
any layered sine--Gordon model, whatever be the mass matrix, can be  
mapped onto an equivalent gas of topological excitations, whose  
interaction potentials are determined by the inversion of a two-dimensional 
propagator of the form $-\partial^2 + \underline{\underline{M^2}}$. 
Any such propagator, upon backtransformation to coordinate space, can  
only lead to a logarithmic behaviour for the vortex interactions at  
small and large distances, and consequently, cannot possibly reproduce  
the confining linear long-range intralayer interaction given in Eq.~(8.42)  
of Ref.~\cite{BlEtAl1994} and in Ref.~\cite{LSGPierson}. The 
candidate~\cite{LSGPierson} for a mass matrix  
$\underline{\varphi}^{\rm T} \, {\underline{\underline m}}^2 
\underline{\varphi} = J\,\sum_{i=1}^{N-1} (\varphi_i - \varphi_{i+1})^2 $ 
has also been discussed in  
Refs.~\cite{JeNaZJ2006,Na2006,NaEtAl2007jpc}. This candidate 
interaction is inspired by a discretization  
of the anisotropic 3D-SG model~\cite{LSG3D},  
but it cannot reproduce the linear  
confining potential needed for the description of the Josephson-coupled  
case~\cite{NaEtAl2007jpc}. The layer-dependent transition temperature of  
this model is $T_{\rm c} \propto N^{-1}$ and decreases with the number of  
layers, and for general $N$, the mass matrix ${\underline{\underline m}}^2$ 
also leads to different short- and long-range intralayer potentials as  
compared to Eq.~(\ref{pot_limit}) and cannot be used for the description  
of magnetically coupled $N$-layer systems, either~\cite{NaEtAl2007jpc}. 
Finally, let us note that a suitable model for the Josephson coupled  
layered system could probably be constructed if the interlayer interaction 
term is  
represented by a compact variable, i.e., one couple the phase (compact)  
fields between the 2D planes~\cite{BeCaGi2007} and not the the dual fields. 
 
% 
% RG Analysis of the Multi-Layer Sine--Gordon Model 
% 
\section{RG Analysis of the Multi-Layer Sine--Gordon Model} 
\label{sec3} 
 
The above statements on the MLSG model are based on the bare action  
where the coupling parameters of the theory are fixed. However, only 
a rigorous RG analysis enables one to construct the phase diagram in  
a reliable manner. 
For $N=2$ layers, the phase structure and the vortex properties of  
the magnetically coupled layered system have already been considered  
with a real space RG approach \cite{CoGeBl2005} using a two-stage  
procedure, and a momentum space RG method \cite{LSGPierson}  
on the basis of the dilute gas approximation has also been used.  
%  
% Fig. 1  
%  
\begin{figure}[htb]  
\begin{center}  
\begin{minipage}{14cm} 
\begin{center} 
\epsfig{file=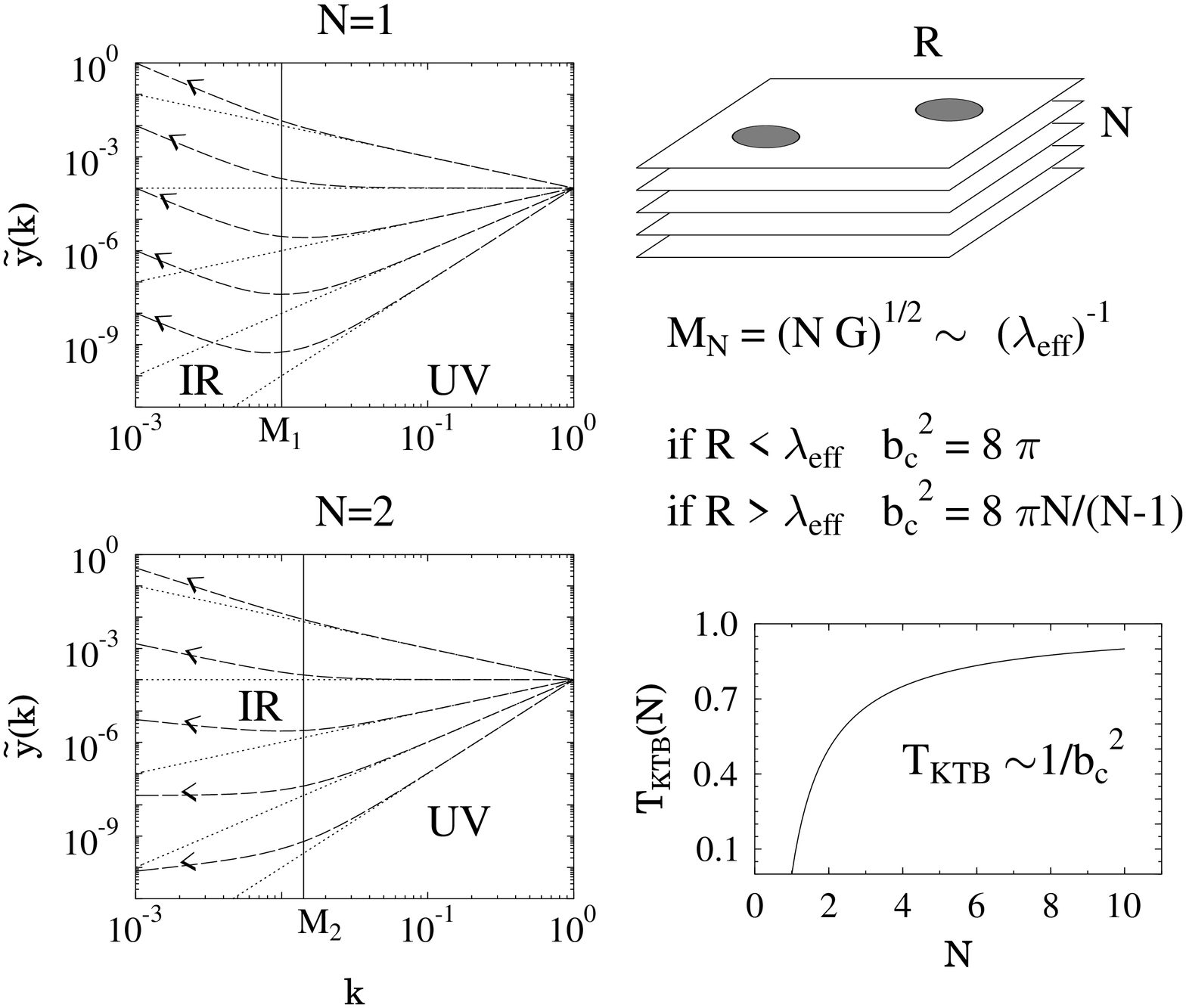,width=0.6\linewidth}  
\caption{\label{fig1}In the left panels, the  
mass-corrected scaling [see Eq.~(\ref{sol})] of the  
dimensionless Fourier amplitude $\tilde y$ of the MLSG 
model for $N=1$ (top) and for $N=2$ (bottom)  
layers is represented graphically for  
$b^2 = 4\pi, 8\pi, 12\pi, 16\pi, 20\pi$ (from top to bottom 
on each panels, see the dashed curves). 
We use $G=0.0001$ in order to  
have the UV and IR regimes conveniently located on the  
plots, which start at the UV scale $\Lambda = 1$. The dotted line  
is the extrapolation of the UV ($k\gg M_N$) scaling to the  
IR ($k\ll M_N$) region.  
For $N=1$ layers, $\tilde y$ is always  
relevant ($\sim k^{-2}$) in the IR. For $N=2$ layers,  
$\tilde y$ is relevant for $b^2 < 16\pi$ in the IR 
and irrelevant for $b^2 > 16\pi$. Thus, the 2-layer MLSG  
model undergoes a KTB type phase transition at $b_c^2 = 16\pi$.  
In general, the KTB transition temperature of the MLSG model  
is layer-dependent $T^{(N)}_{\rm KTB} = (1-N^{-1})T^{\star}_{\rm KTB}$. 
If the system has a finite volume ($R<\infty$), the  
thermodynamic limit cannot be taken automatically and, as a  
simple realization of the finite size effect, a momentum scale  
$k_{\rm min}\sim 1/R$ appears in the model. For $R<\lambda_{\rm eff}$  
(i.e. $k_{\rm min} > \sqrt{N G}=M_N$), the phase structure of the  
MLSG model is determined by the UV scaling which predicts a KTB  
type phase transition at $b_c^2 = 8\pi$ for any number of layers. 
}  
\end{center} 
\end{minipage} 
\end{center}  
\end{figure} 
Here, we apply a generalized multi-layer, 
multi-field Wegner--Houghton (WH) RG   
analysis developed by us for the layered SG type models  
\cite{NaNaSaJe2005,JeNaZJ2006,NaSa2006,Na2006,NaEtAl2007jpc}  
to the MLSG model with an arbitrary  
numbers of layers. In the construction of the WH--RG equation, 
the blocking transformations~\cite{Wi1971} are realized by  
a successive elimination of the field fluctuations in the  
direction of decreasing momenta, in infinitesimal momentum  
shells, about the moving sharp momentum cutoff $k$ 
(see Ref.~\cite{WeHo1973}).  
The physical effects of the eliminated modes are transferred  
to the scale-dependences of the coupling constants [e.g.,  
$y \equiv y(k)$]. The WH-RG equation in the local potential  
approximation (LPA) for the MLSG model with $N$ layers reads 
\begin{equation}  
\label{wh_rg_N}  
(2+k \, \partial_k) \,\, \tilde V_k = - \frac{1}{4\pi}  
\ln \left[ {\mr{det}}%  
\left(\delta_{ij} +  \partial_{\varphi_i}\partial_{\varphi_j}  
\tilde V_k \right) \right],  
\end{equation}  
where we have defined the dimensionless blocked potential as 
$\tilde V_k \equiv k^{-2} \, V_k$. We make the following 
ansatz for the blocked potential, 
\begin{align}  
\label{ansatz}  
\tilde V_k =  
\hf {\tilde G}_{k}  
\left(\sum_{n=1}^N \varphi_n \right)^2 
+ {\tilde U}_{k}(\varphi_1, \cdots \varphi_N),   
\end{align}  
where the scale-dependence is encoded in the dimensionless  
coupling constants $\tilde y(k)$ and $\tilde G(k)$ which are all  
related to their dimensionful (no tilde) counterparts by a  
relative factor $k^{-2}$. Inserting the ansatz (\ref{ansatz})  
into Eq.~(\ref{wh_rg_N}), the right hand side becomes periodic,  
while the left-hand side contains both periodic and non-periodic  
parts~\cite{NaNaSaJe2005,Na2006}.  
 
In order to go beyond the dilute-gas approximation, 
we calculate a mass-corrected  
UV approximation of Eq.~(\ref{wh_rg_N}) 
by expanding the logarithm of  
the determinant in the right hand side of  Eq.~(\ref{wh_rg_N})  
in powers of the periodic part of the blocked potential. 
Because this procedure has been discussed at length in 
Refs.~\cite{NaNaSaJe2005,JeNaZJ2006,Na2006}, 
we immediately state the result [cf.~Eq.~(43) of  
Ref.~\cite{NaNaSaJe2005}], 
\begin{equation}  
\label{sol}  
{\tilde y(k)} = {\tilde y}(\Lambda)   
\left(\frac{k^2 + N \, G}{\Lambda^2 + N \, G}\right)^{\frac{b^2}{N 8\pi}}  
\left(\frac{k}{\Lambda}\right)^{\frac{(N-1)b^2}{N 4\pi} -2},  
\end{equation}  
with the initial value ${\tilde y}(\Lambda)$ at the UV  
cutoff $k = \Lambda$. Let us note that in our RG approach the  
dimensionful $G$ and $b^2$ are scale-independent  
constants. We can immediately read off from Eq.~(\ref{sol})  
the critical value  
$b^2_{c} =8\pi/(1-N^{-1})$ and the corresponding KTB temperature  
$T^{(N)}_{\rm KTB} \sim b^{-2}_c = T^{\star}_{\rm KTB} (1-N^{-1})$.  
The fugacity $\tilde y$ is irrelevant (decreasing) for  
$b^2>b^2_{c}$ and relevant (increasing) for $b^2<b^2_{c}$  
for decreasing scale $k$ (see Fig.~\ref{fig1}). 
Our RG approach provides a consistent scheme to calculate higher  
order corrections to the linearization in 
the periodic part of the blocked potential, which is 
equivalent to higher-order corrections to the dilute-gas 
approximation. For $N=1$, the mass-corrected UV scaling law~(\ref{sol}),  
obtained for the massive SG model, recovers the scaling  
obtained in Refs.~\cite{PiVa2000,IcMu1994} (no phase transition).  
 
% 
% Conclusion and Summary 
% 
\section{Conclusion and Summary} 
\label{sec4} 
 
In conclusion, we propose the multi-layer  
sine--Gordon (MLSG) Lagrangian as a quantum field theoretical model  
for the vortex properties of magnetically coupled              
layered superconductors. Note that the MLSG model cannot be  
assumed to belong to the same universality class as the  
layered Ginzburg--Landau model~\cite{NaEtAl2007jpc},  
which entails a discretization of the Ginzburg--Landau model  
in one of the spatial directions. 
The mapping of the MLSG model onto the gas  
of topological defects is used to clarify the suitability of the  
MLSG model to magnetically coupled layered systems. We investigate  
the scaling laws for the MLSG model using a functional formulation of   
the Wegner-Houghton RG approach in the local potential approximation.   
The linearization of the RG flow in the periodic part of the blocked   
potential (and not in the full potential) enables us to incorporate   
the effect of the interlayer interaction into the mass-corrected   
UV scaling laws, which improve the dilute gas approximation.  
The mass-corrected Wegner--Houghton UV scaling laws indicate  
that for general interlayer interactions of the type of  
Eqs.~(\ref{mass_matrix}), one finds two phases separated by the  
critical value $b_c^2 = 8 \pi/(1 - N^{-1})$, where $N$ is the  
number of layers. This determines the layer-dependence of the KTB  
transition temperature  
$T^{(N)}_{\rm KTB} = T^{\star}_{\rm KTB} \, (1 - N^{-1})$ 
in full agreement with Refs.~\cite{Pu1993,CoGeBl2005}.  
Perhaps, further investigations of the MLSG model (e.g., beyond  
the local potential approximation) and other  
generalizations of the momentum-space RG studies presented here 
could enrich our understanding of the layered structures. 
 
% 
% Acknowledgments 
% 
\section*{Acknowledgments} 
 
The authors acknowledge insightful discussions 
with Professor J. Zinn--Justin and thank Professor  
J. R. Clem for valuable remarks. 
I.N.~would like to acknowledge the kind hospitality of 
Max--Planck--Institut, Heidelberg on the occasion of a number of  
guest researcher appointments, and U.D.J.~acknowledges support from  
the Deutsche Forschungsgemeinschaft with the  
Heisenberg program (contract JE285/3--1). 
S.N. ackowledges  support via the Oveges program of the National
Office for Research and Technology of Hungary and the support by 
the Univesritas Foundation, Debrecen.


\begin{thebibliography}{10} 
 
\bibitem{NaEtAl2007jpc} 
I. N\'andori, U.~D. Jentschura, S. Nagy, K. Sailer, K. Vad, and S. 
  M\'esz\'aros, J. Phys.: Condens. Matter {\bf 19},  236226  (2007). 
 
\bibitem{Pe1964} 
J. Pearl, Appl. Phys. Lett. {\bf 5},  65  (1964). 
 
\bibitem{Pu1993} 
V. Pudikov, Physica C (Amsterdam) {\bf 212},  155  (1993). 
 
\bibitem{BlEtAl1994} 
G. Blatter, M.~V. Feigel'man, V.~B. Geshkenbein, A.~I. Larkin, and V.~M. 
  Vinokur, Rev. Mod. Phys. {\bf 66},  1125  (1994). 
 
\bibitem{CoGeBl2005} 
A. De~Col, V.~B. Geshkenbein, and G. Blatter, Phys. Rev. Lett. {\bf 94}, 
  097001  (2005). 
 
\bibitem{Ef1979} 
K.~B. Efetov, Zh. \'{E}ksp. Teor. Fiz. {\bf 76},  1781  (1979), [JETP {\bf 49}, 
  905 (1979)]. 
 
\bibitem{ArKr1990} 
S.~N. Artemenko and A.~N. Kruglov, Phys. Lett. A {\bf 143},  485  (1990). 
 
\bibitem{BuFe1990} 
A. Buzdin and D. Feinberg, J. Phys. (Paris) {\bf 51},  1971  (1990). 
 
\bibitem{FeGeLa1990} 
M.~V. Feigel'man, V.~B. Geshkenbein, and A.~I. Larkin, Physica C {\bf 167}, 
  177  (1990). 
 
\bibitem{Cl1991} 
J.~R. Clem, Phys. Rev. B {\bf 43},  7837  (1991). 
 
\bibitem{Fi1991} 
K.~H. Fischer, Physica C {\bf 178},  161  (1991). 
 
\bibitem{KoVi1991} 
A.~E. Koshelev and V.~M. Vinokur, Physica C {\bf 173},  465  (1991). 
 
\bibitem{MiKoCl2000} 
R.~G. Mints, V.~G. Kogan, and J.~R. Clem, Phys. Rev. B {\bf 61},  1623  (2000). 
 
\bibitem{dG1966} 
P.~G. de~Gennes, {\em Superconductivity of Metals and Alloys} (Addison-Wesley, 
  Reading, MA, 1966). 
 
\bibitem{ClemPancake} 
J. R. Clem, J. Supercond. Incorp. Novel Magn. {\bf 17}, 613 (2004). 
This paper contains both new results and historical remarks on the  
development of the concept of pancake vortices. 
 
\bibitem{GoHo2005} 
R. Goldin and B. Horovitz, Phys. Rev. B {\bf 72},  024518  (2005). 
 
\bibitem{BeCaGi2007} 
L. Benfatto, C. Castellani, and T. Giamarchi, Phys. Rev. Lett. {\bf 98}, 
  117008  (2007). 
 
\bibitem{BeCaGi2007magnetic} 
L. Benfatto, C. Castellani, and T. Giamarchi,  
  e-print: arXive: 0707.1271 (2007). 
 
\bibitem{Ar2007}  
A. N. Artemov, e-print: arXive: 0708.2775 (2007). 
 
\bibitem{KTBPhase} 
J. M. Kosterlitz, D. J. Thouless, J. Phys. C {\bf 6}, 1181 (1973); V. L. 
  Berezinskii, Sov. Phys. JETP {\bf 32}, 493 (1971). 
 
\bibitem{PiVa2000} 
S.~W. Pierson and O.~T. Valls, Phys. Rev. B {\bf 61},  663  (2000). 
 
\bibitem{ReEtAl1996} 
J.~M. Repaci, C. Kwon, Q. Li, X. Jiang, T. Venkatessan, E.~R. Glover, C.~J. 
  Lobb, and R.~S. Newrock, Phys. Rev. B {\bf 54},  R9674  (1996). 
 
\bibitem{Gi1952} 
V.~L. Ginzburg, Zh. \'{E}ksp. Teor. Fiz. {\bf 23},  236  (1952). 
 
\bibitem{SG2D}  
I. N\'andori {\it et al.}, Phys. Rev. D {\bf 63}, 045022 (2001);   
Phil. Mag. B {\bf 81}, 1615 (2001).  
 
\bibitem{LaDo1971}  
W.~E. Lawrence and S. Doniach, Proc. 12th Int. Conf. on Low Temp. Phys. (ed.~E.  
  Kanda), Kyoto, Japan, 361 (1971).  
 
\bibitem{ChDuGu1995}  
S.~N. Chapman, Q. Du, and M.~D. Gunzburger, SIAM J. Appl. Math. {\bf 55},  156  
  (1995).  
 
\bibitem{RESCALE}  
R. A. Klemm and J. R. Clem, Phys. Rev. B {\bf 21}, 1868 (1980); G. Blatter, V.  
  G. Geshkenbein and A. I. Larkin, Phys. Rev. Lett. {\bf 68}, 875 (1992).  
   
\bibitem{3DXY}  
B. Chattopadhyay and S. R. Shenoy, Phys. Rev. Lett. {\bf 72}, 400 (1994);   
Phys. Rev. B {\bf 51}, 9129 (1995).  
 
\bibitem{Pi1995prb} 
S.~W. Pierson, Phys. Rev. B {\bf 51}, 6663 (1995). 
 
\bibitem{Sa1978}  
R. Savit, Phys. Rev. B {\bf 17}, 1340 (1978);  
  Rev. Mod. Phys. {\bf 52}, 453 (1980);  
 
\bibitem{LSGPierson} 
S. W. Pierson, O. T. Valls, Phys. Rev. B {\bf 45}, 13076 (1992); S. W. Pierson 
  {\it et al.}, Phys. Rev. B {\bf 45}, 13035 (1992). 
 
\bibitem{ZJ1996} 
J. Zinn-Justin, {\em Quantum Field Theory and Critical Phenomena}, 3rd ed. 
  (Clarendon Press, Oxford, 1996). 
 
\bibitem{JeNaZJ2006} 
U.~D. Jentschura, I. N\'{a}ndori, and J. Zinn-Justin, Ann. Phys. (N.Y.) {\bf 
  321},  2647  (2006). 
 
\bibitem{Na2006} 
I. N\'andori, J. Phys. A {\bf 39},  8119  (2006). 
 
\bibitem{LSG3D} 
I. N\'andori {\it et al.}, Phys. Rev. D {\bf 69}, 025004 (2004); J. Phys. G 
  {\bf 28}, 607 (2002). 
 
\bibitem{NaNaSaJe2005} 
I. N\'{a}ndori, S. Nagy, K. Sailer, and U.~D. Jentschura, Nucl. Phys. B {\bf 
  725},  467  (2005). 
 
\bibitem{NaSa2006} 
I. N\'andori and K. Sailer, Phil. Mag. {\bf 86},  2033  (2006). 
 
\bibitem{Wi1971} 
K.~G. Wilson, Phys. Rev. D {\bf 3},  1818  (1971). 
 
\bibitem{WeHo1973} 
F.~J. Wegner and A. Houghton, Phys. Rev. A {\bf 8},  401  (1973). 
 
\bibitem{IcMu1994} 
I. Ichinose and H. Mukaida, Int. J. Mod. Phys. A {\bf 9},  1043  (1994). 
 
\end{thebibliography}
\end{document}